\documentclass[pdflatex,sn-mathphys]{sn-jnl}% Math and Physical Sciences Reference Style
\usepackage{xurl}

\theoremstyle{thmstyleone}%
\theoremstyle{thmstyletwo}%

\theoremstyle{thmstylethree}%
\newtheorem{definition}{Definition}%

\begin{document}

\title{\bf{Ruliology: Linking Computation, Observers and Physical Law}}
%Dean, Hatem - tentative title. feel free to modify


\author[1]{\fnm{  \, \, \, \, \, \, \, \,  Dean} 
\sur{Rickles}}\email{dean.rickles@sydney.edu.au}
\equalcont{All authors contributed equally to this work.}
\author[1,2,4]{\fnm{Hatem} \sur{Elshatlawy}}\email{hatem@wolfram.com}
\equalcont{All authors contributed equally to this work.}
\author[3,4]{\fnm{ \, \, \, \, \, \, \, \, \,  Xerxes D.} 
\sur{Arsiwalla}}\email{x.d.arsiwalla@gmail.com}
\equalcont{All authors contributed equally to this work.}

% XDA = feel free to adjust order of authors as required 

\affil[1]{\orgname{University of Sydney}, \orgaddress{\city{Sydney}, \country{Australia}}}
\affil[2]{\orgname{RWTH Aachen University}, \orgaddress{\city{Aachen}, \country{Germany}}}
\affil[3]{\orgname{Pompeu Fabra University}, \orgaddress{\city{Barcelona}, \country{Spain}}}
\affil[4]{\orgname{Wolfram Research}, \orgaddress{\country{USA}}}


\abstract{Stephen Wolfram has recently outlined an unorthodox, multi-computational approach to fundamental theory, encompassing not only physics but also mathematics in a structure he calls ``The Ruliad,''  understood to be the entangled limit of all possible computations. In this framework, physical laws arise from the sampling of the Ruliad by observers (including us). This naturally leads to several conceptual issues, such as what kind of object is the Ruliad? What is the nature of the observers carrying out the sampling, and how do they relate to the Ruliad itself? What is the precise nature of the sampling? This paper provides a philosophical examination of these questions, and other related foundational issues, including the identification of a limitation that must face any attempt to describe or model reality in such a way that the modeller-observers are included.}

%%================================%%
%% Sample for structured abstract %%
%%================================%%




\maketitle

%\clearpage
%\tableofcontents

\section{Introduction: Foundations of Physical Theory}\label{sec1}

\begin{quote}
   \noindent  If we had invented the digital computer before inventing graph paper, we might have a very different theory of the universe today.\\ Jacques Vallee, \emph{Dimensions}\footnote{\emph{Dimensions: A Casebook of Alien Contact} (Anomalist Books, 2008, p. 287).}
\end{quote}

\vspace{5pt}

 \noindent Often when thinking about the modeling of reality in physical theories, we employ an abstract space that is supposed to represent all possible states of a system, a modal arena, one point of which will correspond to the present state of a `real-world' system.  This provides the kinematical structure of a theory when forces are ignored (yielding a larger space of possibilities than is physically allowed) and the dynamics when forces are included (yielding the so-called nomologically posssible states). In general, what is not in the space is not a possibility; and what is not a possibility is not in the space. This can be a \emph{universal} state space  as in the geometrodynamics of John Wheeler, where it is known as ``superspace'' \cite{wheegrav}. Here, `points' of the space are 3-dimensional geometric configurations (i.e. Riemannian geometries on a 3-manifold) of the universe and histories are then represented as trajectories (paths through the space, generating spatiotemporal worlds), corresponding to possible universes---the space of 3-geometries is understood as the space of 3-metrics `quotiented' by the diffeomorphism group (the invariance group of general relativity), identifying  those metrics differing by elements of that group.\footnote{This is formally akin to the notion of moduli spaces in algebraic geometry, where points in the moduli space correspond to isomorphism classes of algebro-geometric objects and trajectories yield a formal notion of dynamics on the space. Moduli spaces are useful for classification problems, where coordinatizing the space is useful for studying various classes of deformations (of the moduli parameters) corresponding to the objects in question (see \cite{mod} for a useful introduction).} Furthermore, in quantum geometrodynamics, we envisage a wave-function over this configuration space which assigns amplitudes for the various types of state of the universe.\footnote{This is the infamous Wheeler-DeWitt equation with its problematic interpretation in terms of dynamical evolution resulting from the absence of a time-parameter $t$, itself stemming from the fact that time evolution is a kind of diffeomorphism in general relativity and so belonging in the category of gauge degrees of freedom rather than the structure to be assigned a physical interpretation (see \cite{rick1}).} While presented as a rather fundamental description of physics, even the superspace point of view clearly stands several rungs up on the ontological ladder, presupposing several layers of deeper structure.\footnote{This holds also for the `upgrade' of quantum geometrodynamics from 3-geometries and their histories to spin-networks and their spin-foams, though the latter formalism features a slightly more primitive structure in that it involves abstract graphs as its fundamental objects and whose relations build up the various layers of structure we associate with out physical theories.}

The ``Physics Project'' recently initiated by Stephen Wolfram \cite{nks,w1,w2,w3,swcon} aims to describe how all other levels of structure are built from the ground up, that is, from ontological ground zero. The basic structure is not a set of elements as such, but a \emph{totality} that can then be decomposed to generate possible universes, including our own. This structure is called the ``Ruliad'' or ``Rulial space'' and is usually expressed informally as the result of carrying out a process (or, rather, many such) to infinity, yielding ``the entangled limit of all possible computations'': it is what is generated by carrying out all possible rules in every possible way \cite{ruliad,swtm}. It is computationally exhaustive. Like the universal state space of geometrodynamics or moduli spaces, sections of the Ruliad correspond to possible histories of universe (though unlike the former two, the Ruliad is a purely syntactic structure, defined independently of any \emph{a priori} geometric notions).

\emph{Ruliology}, a term coined by Stephen Wolfram, delves into the intricacies of rule space and how different rules can lead to diverse and complex behaviors. It represents the study of the Ruliad, a profound and encompassing framework within the Physics Project that serves as the theoretical foundation for understanding the myriad of computational universes. Instead of treating reality as a mere collection of isolated entities, Ruliology embraces the idea of the Ruliad — a vast, interconnected web of all conceivable computations, executed through every possible rule. This intricate and boundless space is not just a speculative novelty; it provides a comprehensive map from which individual universes, including ours, can be derived. Such a perspective challenges traditional views on the nature of reality and paves the way for a more unifying, computational understanding of the cosmos. Given its ambitious scope, the potential depth of insight into the nature of rules, and its profound implications, Ruliology demands rigorous exploration and merits earnest attention in the broader scientific discourse.

It is perhaps worth remarking up front on the similarities to David Deutsch's notion of a ``constructor'' here \cite{constructor}, because in that case, as in Wolfram's approach, one is demonstrating existence through a constructive (computationally conceived) procedure---both also find some insufficiency in the orthodox Turing machine model of a universal computer as a model of reality. What is possible can be constructed \emph{physically} from some rule (or ``task'' in Deutsch's terminology), and what cannot is impossible (i.e. there is no such constructor up to the task). Note, also, that as constructive theories, they have an end-goal in mind (namely, that which is to be constructed), and so both contain teleological elements.\footnote{Indeed, Deutsch has compared his own approach to cybernetics (which involves steering systems to pre-defined goals), which he describes as a possible ``early avatar'' (\url{https://www.edge.org/conversation/constructor-theory}). But while Deutsch believes he has provided a perfectly non-abstract, physical description of the world (and, e.g. talks of abstract computers not making sense), he then slips to mentioning information as if it itself is concrete. It is not, and this is where Wolfram's approach has the advantage, since it directly brings in the additional element that allows for the inclusion of information into the model. If constructor theory has cybernetics as an early avatar, then 2nd-order cybernetics is an early avatar of the Wolfram model. Indeed, Deutsch's claim about materiality being necessary for realization is a tautology since by ``realization'' he \emph{means} within a material system. This ignores the fact that there is clearly an information template beyond that realization which is what such realizations in matter are realizing. There is quite simply no getting around the fact that if one adopts an information-based ontology, then one is stepping somewhat outside of orthodox materialism.} The key idea of constructor theory is, then, simply that the focus of fundamental theory should be which transformations of some medium or substrate into another such state can be caused to occur, as well, by implication, as those which cannot be so caused. Given substrate independence, the focus becomes the \emph{transformations} themselves as the ontological core of the theory. The precise nature of this, essentially, modal structure consisting of counterfactuals has yet to have been adequately nailed down, since while the transformations themselves are always grounded in some physical substrate, the counterfactuals, as non-actual by definition, are clearly not (though see \cite{marlettobook} for a discussion of some of the options and problems). 

Other related constructivist approaches which lend similar precedence to processes over substrates include ``Assembly Theory,'' ``Process Theories'', and ``Intuitionistic Physics''. The first of these  \cite{cronin}, focused on the detection of life, is based on \emph{rules of assembly} which take into account the number of independent parts and their connections, such that as the number increases, the need for memory increases which enables the reconstruction of the whole from locally stored rules. Process theories are founded in the framework of category theory and seek to formalize physical operations in diagrammatic terms in which the diagrams (representing morphisms between objects) are expressed as objects and transformations within an appropriate monoidal category  \cite{cqm1,coeckevol,zx0,zx1,zx2,zx3}.  The third approach (mentioned above) seeks a formalization of physical observations and measurements  based on intuitionistic logic, rather than classical logic, to escape, for one, the fact that a physics based on real numbers will face the problem that we will never be able to grasp them, requiring as they do infinite Shannon information to specify their non-repeating decimal expansions \cite{gisin,topos1} (note that intuitionistic mathematics involves a temporal, step-wise process, rather than an eternal, Platonic structure, and this will be important for our later claims about the essential limitations of the Ruliad \emph{qua} fundamental theory). 
% relevant references needed

%Baez, J., & Lauda, A. (2011). A Prehistory of n-Categorical Physics. In H. Halvorson (Ed.), Deep Beauty: Understanding the Quantum World through Mathematical Innovation (pp. 13-128). Cambridge: Cambridge University Press.

The substrate-independent approach described above, in constructor theory, is more or less what Einstein once called the ``principle theory'' method (see, e.g. \cite{einstein}). Rather than dealing with what things are made of, in terms of composition, the method looks at the higher-level principles that any and all things must obey, regardless of their physical constitution.\footnote{But note, again, that on Deutsch's own interpretation, \emph{some} physical substrate or other is required. The substrate would ground particular instances, allowing for task realization.} This transcends particular physical theories and provides a theory of theories: a meta-theory. Wolfram's approach shares this feature of being a theory of theories---one might call it a theory of \emph{all} theories. Among other things, such a meta-theory bears the burden of having to explain how physical notions of space, time, matter, laws, and observers arise, which we identify as the core elements of physical theories. Wheeler recognized this challenge in the 1970s and coined the term ``pregeometry'' in part to address these very issues\footnote{A recent formalization of pregeometry based on homotopy type theory, as well as one based on pre-quantum structures from noncommutative operator algebras can be found in these works: \cite{cpost,opalg,paper2,pregeo,nfold}. These studies borrow from ongoing advances at the foundations of mathematics \cite{hott1,hott2,hott3,hott4} as well as applications of higher category theory to physics \cite{baez,sch1,sch2,sch3}. }.  Note, also, that  in delving to this deeper level, one can evade the usual problems with treating quantum mechanics as a universal theory, in which one can model not just microscopic systems, but also the very agents using quantum mechanics \cite{frauch18}. In the Wolfram  case at least, quantum mechanics is a feature of the structure itself (the so-called multiway description), a consequence, rather than the structure being explicitly constructed to capture quantum features from the outset---the same can be said of the other theories that fall out of the structure without being inserted \emph{ad hoc}.

In this paper we attempt to explain the Wolfram model, focusing more on conceptual issues than formal details, and aim to bring out the role of observers as it appears in the model, showing how it is essential for making sense of standard physics as well as mathematics. The relationship between the Ruliad and the observer is also explicated. We use the basic idea of second-order cybernetics to elucidate a deeper understanding of how the Wolfram model includes the observer, explaining the very generation of the world we see as a kind of observer-selection effect, much like a reference frame in relativistic theories. This allows us, moreover, to provide an account of the nature of ``physical laws'' as sampling-invariance. However, the Wolfram model also lets us see in a new light a fundamental limitation of trying to gain fundamental knowledge of the world from the standpoint of the observer doing the modelling. Let us begin with an account of the basic elements of the Wolfram model.

%  A lot of what one has to say about observers, concerns a new science of subjectivity. And that is also, to some extent, what second-order cybernetics originally wanted to achieve in the context of animal behavior. One can also introduce the observer by laying out the lessons learned from the constructivist philosophy (or perhaps bridging the the above two) 

\section{The Wolfram Model as an Abstract Rewriting System and The Concept of the Ruliad}
\label{s:model}

\begin{quote}
    ``When you come to a fork in the road, take it!''\\
    Yogi Berra
\end{quote}

\begin{figure}[ht]
\centering
\includegraphics[width=0.595\textwidth]{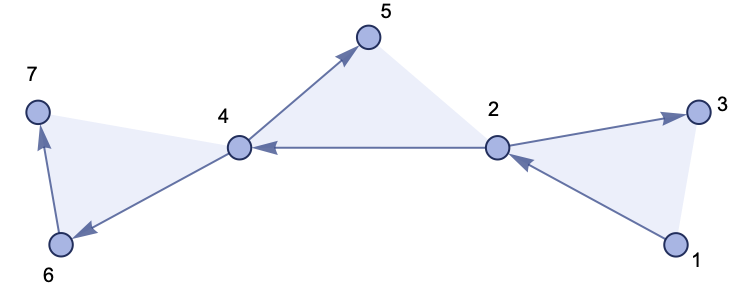}
\caption{An example of a Wolfram model hypergraph with directed hyperedges: $\{\{1,2,3\},\{2,4,5\},\{4,6,7\}\}$.}
\label{fig:Figure1}
\end{figure}

\noindent This section provides a literature overview of the basic abstract rewriting constructions that constitute the Wolfram model, with a particular emphasis on their non-deterministic aspects, as captured via \textit{multiway systems}.\footnote{A more detailed version of this material can be found in \cite{w1,ruliad,mcomp,zx1}, from which portions of the text in this section are taken (as indicated by specific citations).} We discuss the concept of the Ruliad, representing the entangled limit of everything that is computationally possible, which emerges as a key theoretical construct associated to the Wolfram model \cite{ruliad,swtm}.  We begin with a preliminary description of the Wolfram model in terms of diagrammatic rewriting rules acting on hypergraphs. The Wolfram model represents a discrete framework that posits structures such as continuous spacetime geometries which may potentially emerge from large-scale limits of the underlying discrete structures (\cite{nks,w1,zx1,pregeo,graphc}). Furthermore, the evolution of these structures is dictated by various forms of rewriting rules, such as those based on graphs, hypergraphs, or strings\footnote{See \cite{ggrammar,rewrite} for a background overview on  rewriting systems; and \cite{arity1,arity2,arity3} for an overview on hypergraph algebras.}.  To illustrate this, a Wolfram model hypergraph can be represented abstractly as a finite collection of ordered or unordered relations (hyperedges) between labelled nodes, as defined below\footnote{For more detail, see \cite{zx1}, section 2, from which comes some part of the material that follows, including Definitions 1 and 2.} and shown in
Figure \ref{fig:Figure1}:

\begin{definition}
A Wolfram Model Hypergraph $H=(V, E)$ is characterized by a finite set of hyperedges $E$ that belong to the non-empty subset of the power set of $V$, i.e., $E \subset \mathcal{P}(V) \backslash\{\emptyset\}$. A  hyperedge in $E$ is an unordered collection of nodes from the vertex set $V$. Hyperedges in the Wolfram model, can be both, either directed or undirected.
\label{def1}
\end{definition}

\noindent One can then define the dynamics of a Wolfram model system in terms of hypergraph rewriting  rules as follows:

\begin{definition}
A `Rewriting Rule,' denoted as $R$, for a spatial hypergraph $H=(V, E)$, is an abstract rewriting rule expressed in the form $\mathrm{H}_1 \rightarrow \mathrm{H}_2$. In this rule, a subhypergraph that matches the pattern $H_1$ is replaced by a subhypergraph that matches the pattern $H_2$.
\label{def2}
\end{definition}

\begin{definition}
A Wolfram model is an abstract rewriting system founded on the principles outlined in definitions \ref{def1} and \ref{def2}. It's worth noting that Wolfram models are not solely limited to hypergraph rewriting systems; they encompass a range of other rewriting systems, including but not limited to string rewriting systems, term rewriting systems (TRS), (hyper)graph rewriting systems, and cellular automata.
\end{definition}

\begin{figure}[h]
\centering
\includegraphics[width=0.55\textwidth]{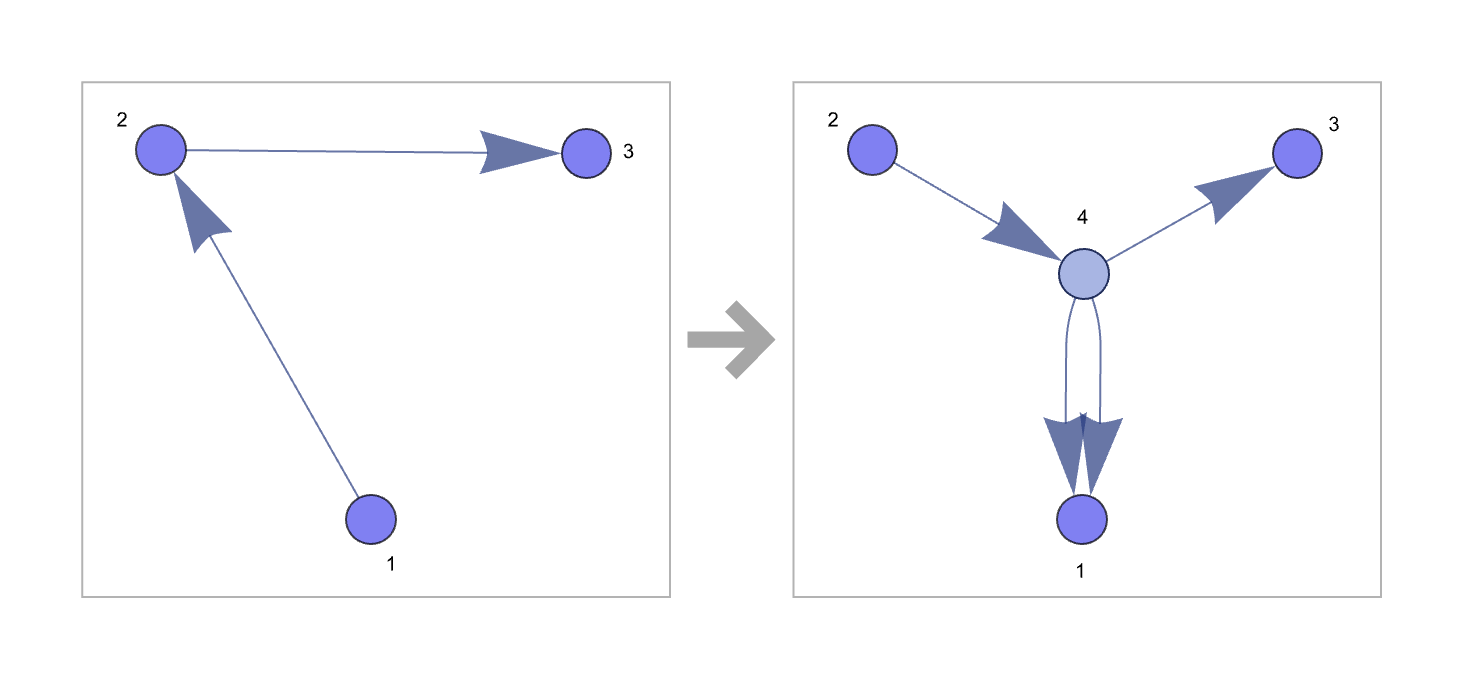}
\caption{A hypergraph transformation rule corresponding to the set-substitution system $\{\{1, 2\},\{2, 3\}\} \rightarrow\{\{4, 1\},\{4, 1\},\{4, 3\},\{2, 4\}\}$.}
\label{fig:Figure2}
\end{figure}

\noindent Every rewriting rule in this context can be formally mapped to a set-substitution system, where a specific subset of ordered (unordered) relations that matches a given pattern is replaced with another distinct subset of ordered (unordered) relations that also corresponds to a particular pattern, as shown in Figure \ref{fig:Figure2}.

%{{{1, 2}, {2, 3}} -> {{4, 1}, {4, 1}, {4, 3}, {2, 4}}}

\begin{figure}[ht]
\centering
\includegraphics[width=0.995\textwidth]{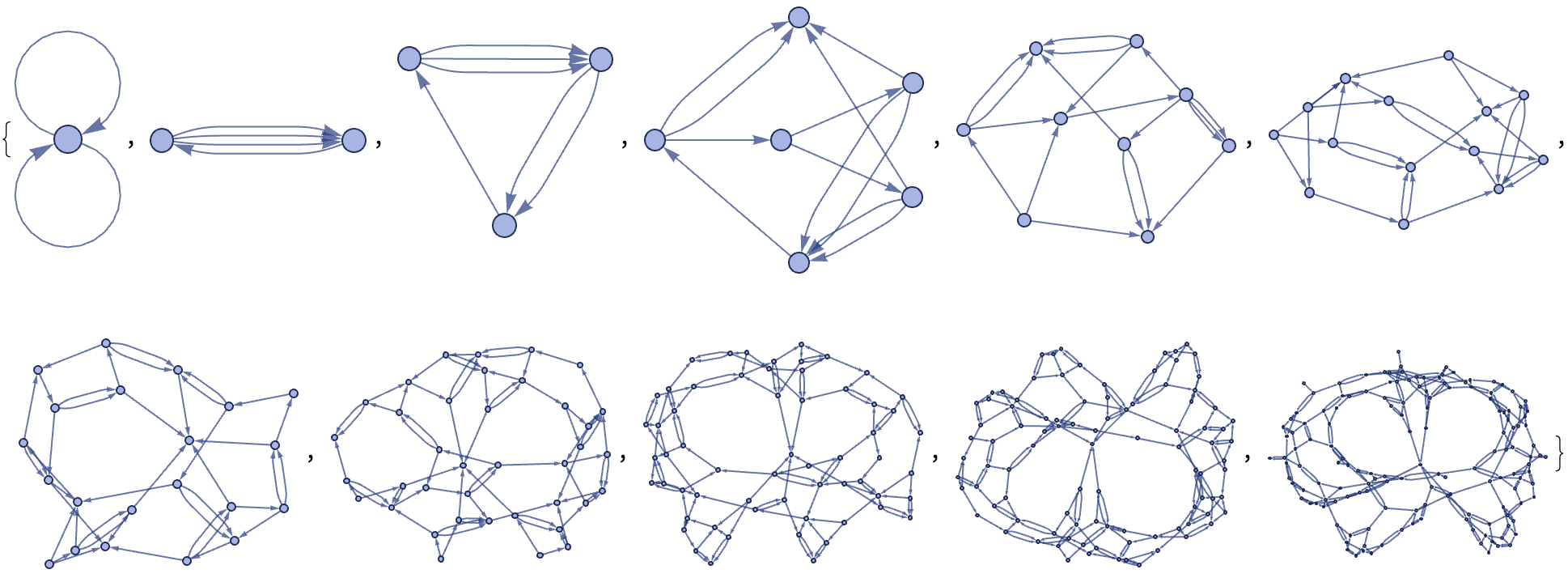}
\caption{The results of the first 10 steps in the evolution history of the set substitution system $\{\{1, 2\},\{2, 3\}\} \rightarrow\{\{4, 1\},\{4, 1\},\{4, 3\},\{2, 4\}\}$, starting from a double self-loop initial condition $\{\{1, 1\},\{1, 1\}\}$.}
\label{fig:Figure3}
\end{figure}

\begin{figure}[ht]
\centering
\includegraphics[width=0.795\textwidth]{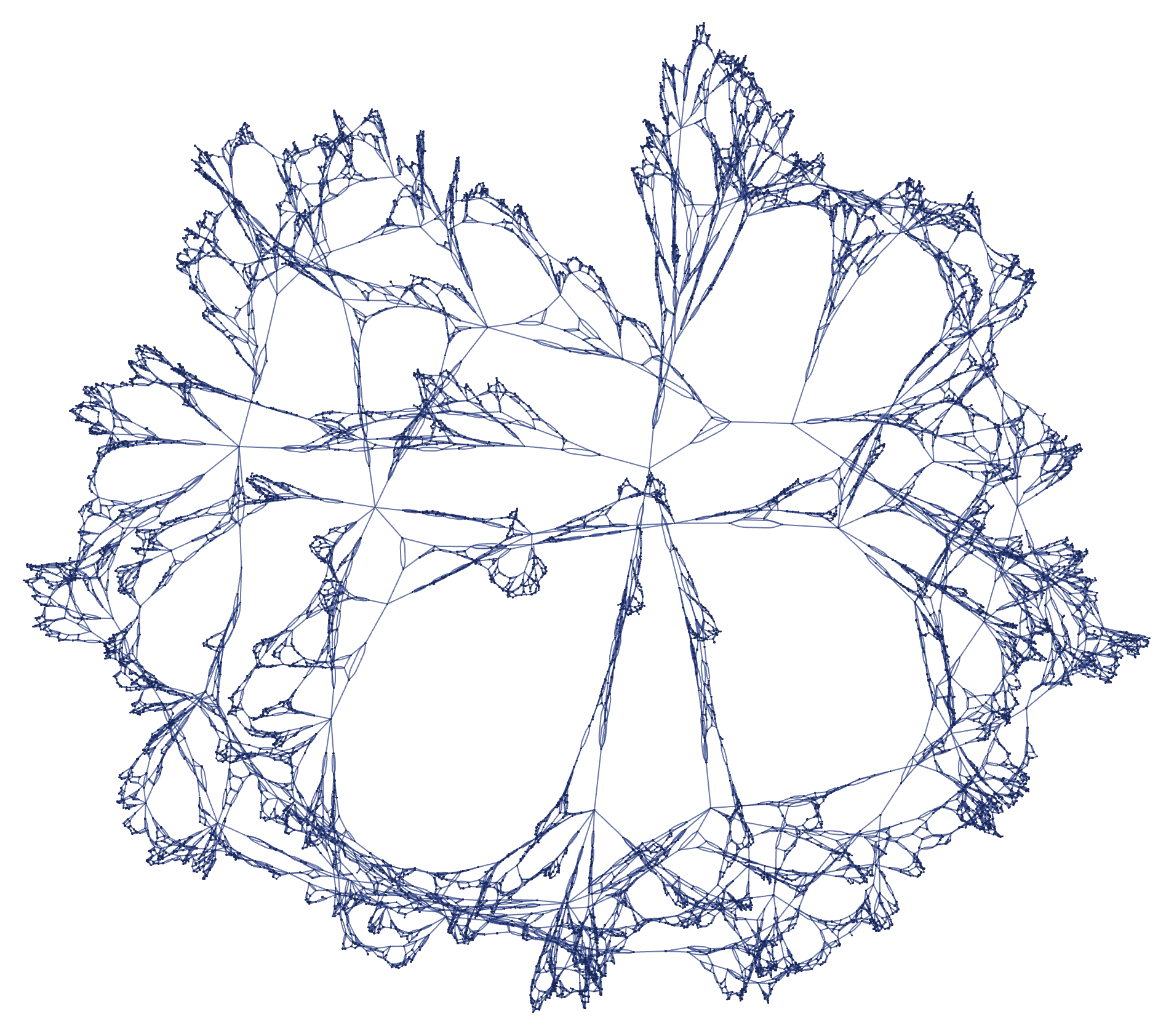}
\caption{ The result after 19 steps of evolution of the set substitution system $\{\{1, 2\},\{2, 3\}\} \rightarrow\{\{4, 1\},\{4, 1\},\{4, 3\},\{2, 4\}\}$, starting from a double self-loop initial condition $\{\{1, 1\},\{1, 1\}\}$.}
\label{fig:Figure4}
\end{figure}

\noindent It is worth noting that the sequence in which transformation rules are applied is generally not predetermined. Even in the simplest scenario where the rule is applied to every matching and distinct subhypergraph (see Figures \ref{fig:Figure3} and \ref{fig:Figure4}), the initial selection of which subhypergraph to transform first remains open-ended. This multiplicity of choices typically leads to different, non-equivalent sequences of evolving hypergraphs. Hence, the evolution of a given spatial hypergraph is inherently non-deterministic due to the absence of a fixed updating order (or rather the possibility of multiple possible updating orders). Therefore, we can treat the Wolfram model as a non-deterministic abstract rewriting system.

\begin{figure}[ht]
\centering
\includegraphics[width=0.495\textwidth]{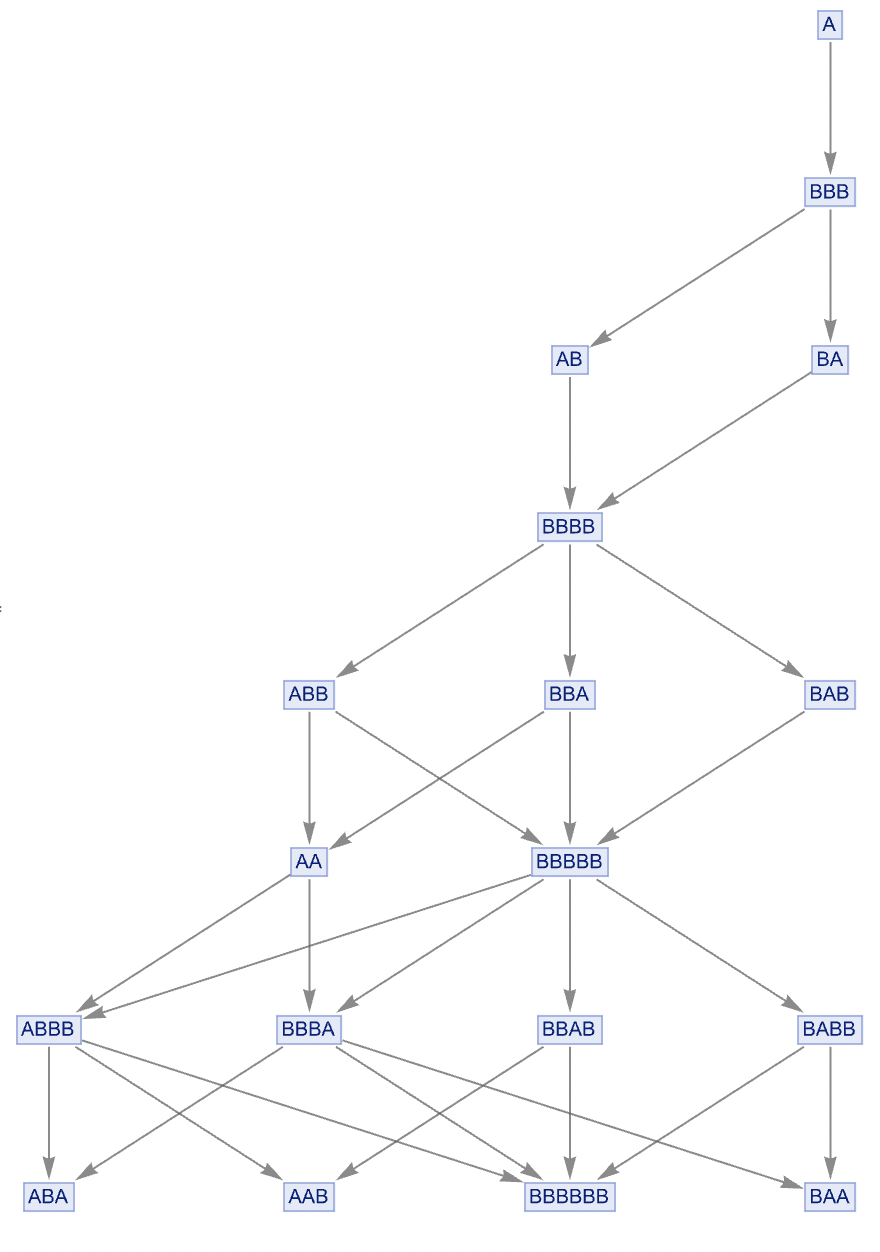}
\caption{The multiway evolution graph corresponding to the first 7 steps in the non-deterministic evolution history of the string rewrite rules: $A \rightarrow BBB, BB \rightarrow A$ (cf. \cite[p. 8]{pregeo}).}
\label{fig:Figure5}
\end{figure}

\noindent Furthermore, within the conventional computational paradigm, systems typically evolve through a series of sequential steps by applying specific rules. However, in the case of Wolfram models (for certain rules), determinism is not inherent. Multiple choices of substitutions are possible, resulting in diverse outcomes. Usually, we select one possibility (e.g. from the possibility space mentioned earlier) and disregard the others (as ways the system \emph{might} have evolved), but the concept of a multiway system allows for simultaneous exploration of all potential choices. A key idea is to consider all those possible threads of history---and to represent these in a single object that we call a multiway graph.\footnote{In some ways this is similar to the universal wave-function of Everettian quantum mechanics \cite{wal2}, though it sits at a far lower-level of structure. Indeed, it sits at what might be called a ``sub-structural'' level (see \cite{paper2} for more on this, including the idea of a `structureless structure' from which structure is generated).} Consider as an example (Figure \ref{fig:Figure5}) a system defined by the string rewrite rules: $A \rightarrow BBB, BB \rightarrow A$. Starting from $A$, the next state has to be $BBB$.  But now there are two possible ways to apply the rules, one generating $AB$ and the other $BA$ (thus forming a fork in the graph).  And if we trace both possibilities we get what we call a multiway system---whose behavior we can represent using a multiway graph. And it's not really difficult to construct multiway system models.  There are multiway Turing machines.  There are multiway systems based on rewriting not only strings, but also trees, graphs or hypergraphs.  There are also multiway systems based on numbers. And all kind of multiway systems. Combinatorially, a multiway system is simply a directed, acyclic graph of states, determined by abstract rewriting rules that inductively generate a (potentially infinite) \textit{multiway evolution graph}, together with a partial order on the rewrite rule applications, determined by their causal structure. 

\begin{figure}[ht]
\centering
\includegraphics[width=0.895\textwidth]{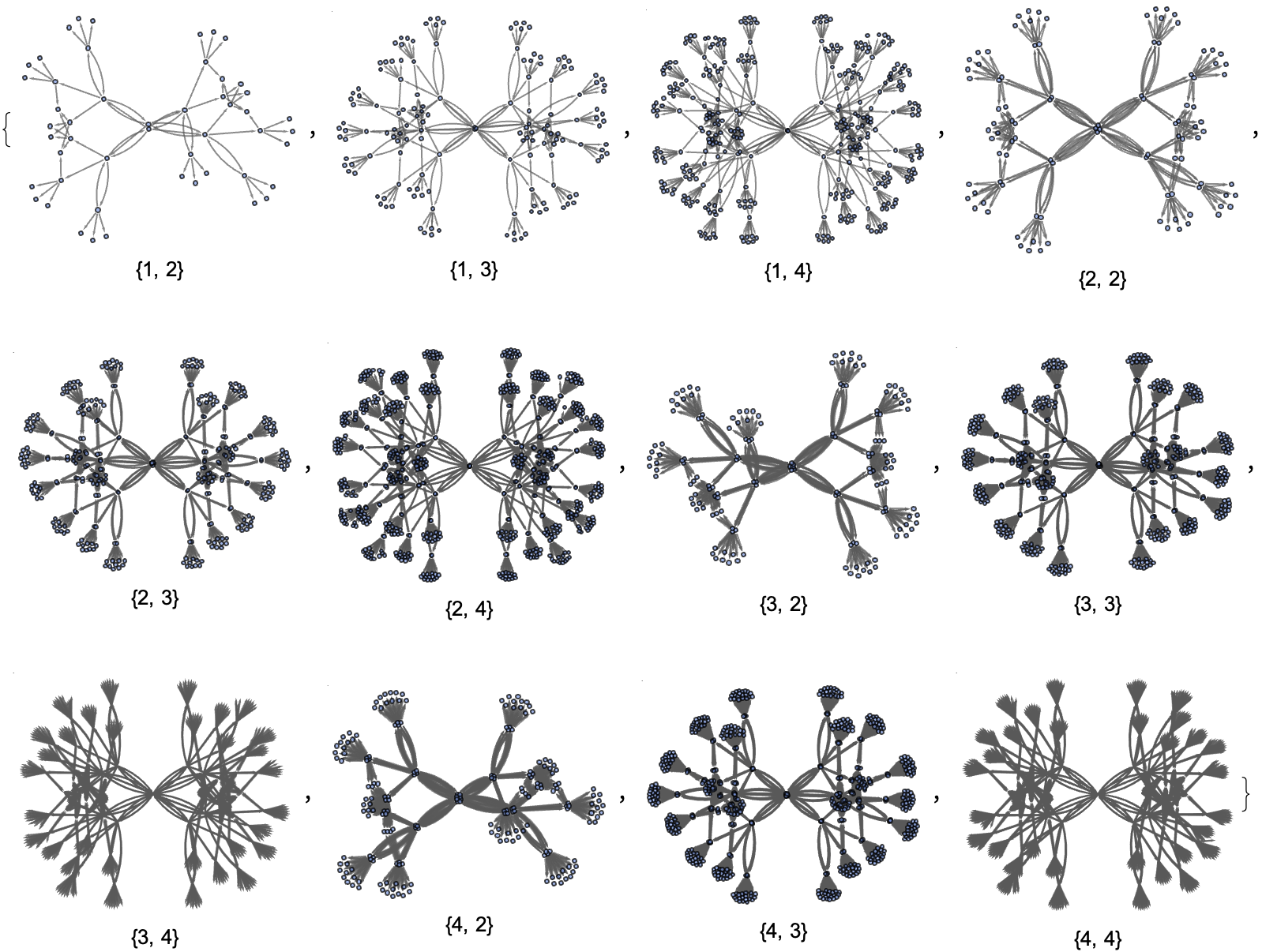}
\caption{Rulial multiway graphs after 3 steps for Turing machines with various numbers of states and colors $\{s,k\}$ (Adapted from \cite{swtm}).}
\label{fig:Figure6}
\end{figure}

\begin{figure}[ht]
\centering
\includegraphics[width=0.895\textwidth]{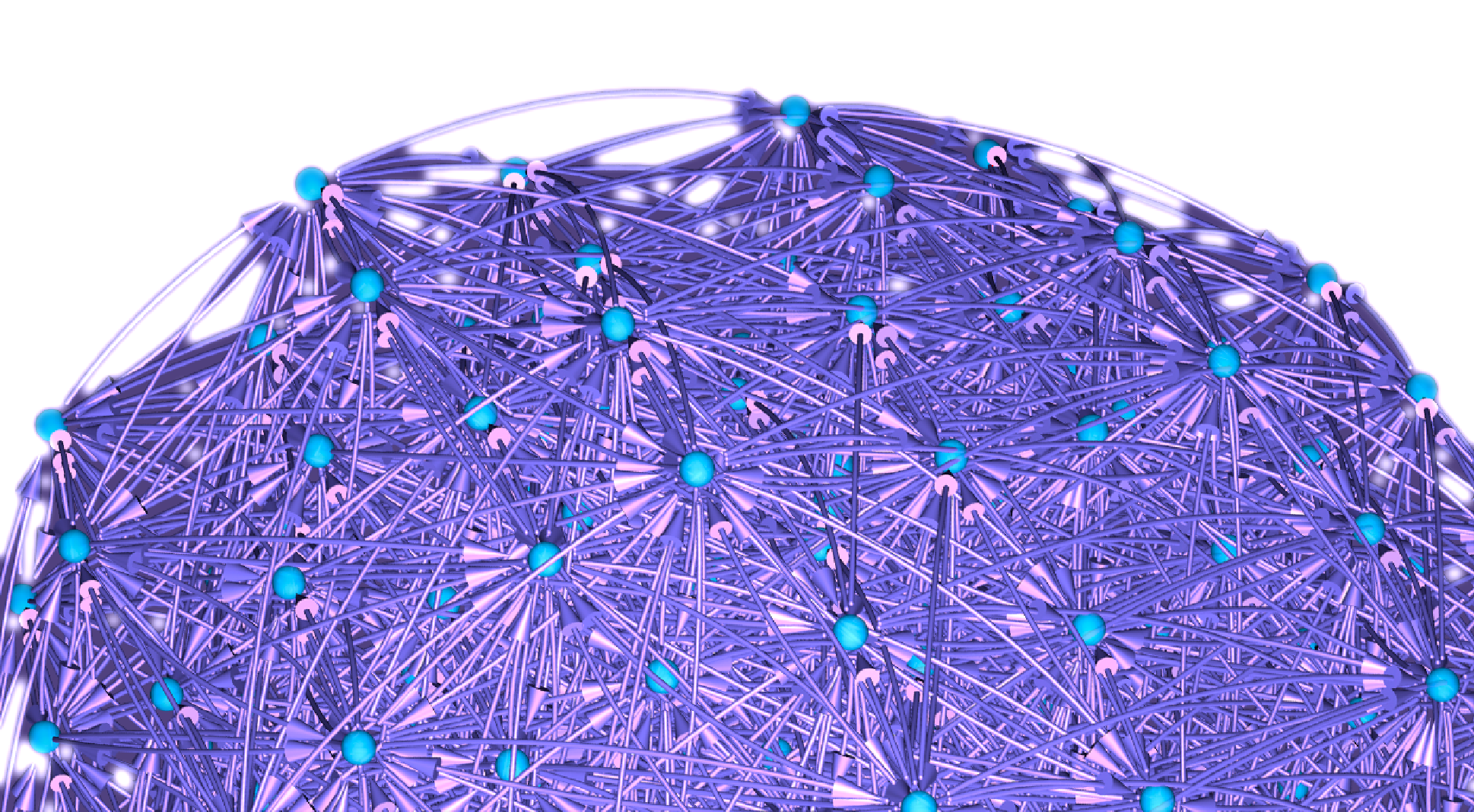}
\caption{A schematic figure of the Ruliad (Adapted from \cite{ruliad}). The image presented can be considered as a rough, finite approximation of the Ruliad. The complete Ruliad encompasses the exploration of infinite limits across all conceivable rules, initial conditions, and steps. Here the nodes and edges are not specific to any single entity. Nodes can represent various entities such as hypergraphs, strings, or states, while the edges signify the myriad potential causal connections between those entities.}
\label{fig:Figure7}
\end{figure}

Now instead of looking at all possible ways a given rule can update these rewrite systems, imagine the structure of spaces created by applying all possible rules. Instead of just forming a multiway graph in which we do all possible updates with a given rule, we form a \textit{rulial} multiway graph in which we follow not only all possible updates, but also all possible rules (an illustrative figure of rulial multiway graphs for Turing machines with various numbers of states and colors $\{s,k\}$ is shown in Figure \ref{fig:Figure6}) \cite{ruliad,swtm}.  This construction allows us to think about the notion of rulial space, i.e. the space of all possible rewriting rules of a given signature. By applying all possible rules in all possible updates we get what we call the \textit{Ruliad} the result of following all possible computational rules in all possible ways (a schematic depiction of a finite approximation of the Ruliad is shown in Figure \ref{fig:Figure7}).\footnote{Here we see another crucial difference to other ensemble theories, such as the Everettian multiverse, which means that such approaches will be automatically subsumed in the Ruliad, as the exhaustive application of just one rule (or category of rules). The Ruliad is instead the ultimate ensemble theory, or the ultimate multiverse.} It’s the ultimate limit of all rulial multiway systems. And as such, it traces out the entangled consequences of progressively applying all possible computational rules. The concept here is to use not just all rules of a given form, but all possible rules. And to apply these rules to all possible initial conditions. And to run the rules for an infinite number of steps. Essentially, the Ruliad involves taking the infinite limits of all possible rules, all possible initial conditions and all possible steps\footnote{A category theoretic description of the limiting rulial multiway system in terms of infinity-categories can be found in \cite{pregeo}.}. Consequently, the Ruliad is in effect a representation of all possible computations. A conceptual  definition of the Ruliad is given below: 

\begin{definition}
\emph{Ruliad}: A meta-structural domain that encompasses every possible rule-based system, or computational  eventuality, that can describe any universe or mathematical structure. It acts as a theoretical space wherein the boundaries between map and territory blur, pushing beyond mere perception and functioning as the ground for the possibility of multi-computation. Within the Ruliad, every conceivable physical and mathematical system can be situated, but their accessibility or meaningfulness is determined by the specific observer-related frames or constructs. The Ruliad is thus a pre-physical framework, and its utilization in physics is to pinpoint the exact rule-based system  that corresponds to our observed reality.
%\end{definition}

\vspace{0.2cm}
\noindent A computational definition is as follows: 
\vspace{0.2cm}

%\begin{definition}
\noindent Let $\mathcal{R}$ be the space of all possible computational (or rewriting) rules. We refer to  $\mathcal{R}$ as the ``Rulial space''.  Consider $r \subseteq \mathcal{R}$. $r$ may represent either a single rule or a collection of rules capable of generating a computational universe $U_r$ (the outcome of all possible computations following a given rule set). The Ruliad, denoted by $\mathbf{R}$, is the collection of all computational universes. That is, $\mathbf{R} = \left\{U_r \mid r \in \mathcal{R}\right\}$. 

\vspace{0.2cm}

\noindent Furthermore, 
\vspace{0.2cm}

\begin{enumerate}
        \item An observer $O$ with a frame or construct $F_O$ can interpret or access a subset of $\mathcal{R}$ based on their specific observational constraints. The observability of any $U_r$ is contingent on $F_O$.
    \item The entirety of $\mathbf{R}$ serves as the foundational ground for multicomputation, transcending the dichotomy between representation and reality.
\end{enumerate} 
\end{definition}

\noindent Ruliology, intricately intersects with this multicomputational paradigm. This nexus between Ruliology and multicomputation sets a fresh foundation for understanding the vastness and versatility of rule space, where different rules can lead to multifaceted and complex behaviors. Thus, one can contend that the Wolfram model encompasses a novel paradigm that extends beyond traditional computation \cite{mcomp}. This is what Stephen Wolfram calls the multicomputational paradigm \cite{mcomp}. It not only traverses the boundaries of physics but also paves the way for a foundational and versatile methodology for crafting models in theoretical science. Historically, three paradigms have dominated theoretical science: mathematical equations, which rely on formulas to describe phenomena; mechanistic models, which provide detailed, step-by-step explanations of how systems function, likened to ``machines'' with distinct parts and processes; and computational models, which view systems as computational entities, allowing for the definition of rules and initial conditions, and then observing the resulting behaviors. However, the multicomputational paradigm goes a step further. It's not just about analyzing specific historical paths but delving into the evolution of all conceivable histories, epitomized by the Ruliad. In many instances, it may not offer insights into specific histories. Instead, what it will describe is what an observer sampling the whole multicomputational process will perceive. This pivotal intersection of Ruliology and multicomputation is where our exploration now focuses.

%Accordingly, one can contend that the Wolfram model encompasses a novel paradigm that surpasses the realm of computation. It introduces a fourth paradigm that we call the \emph{multicomputational paradigm}, which extends beyond physics and serves as a fundamental and versatile methodology for constructing models in theoretical science. The other three paradigms that have historically been used in theoretical science are \emph{mathematical equations}: this is the traditional approach where phenomena are described using mathematical formulas and equations. \emph{mechanistic models}: these are detailed, often step-by-step, models that describe how a system operates. They are like ''machines'' with specific parts and operations that produce an outcome. \emph{computational models}: this paradigm treats systems as computational processes. Instead of trying to describe systems with equations, one defines rules and initial conditions and then ''runs'' the system to see how it behaves. And it's important to emphasize that the multicomputational paradigm is at its core not about particular histories, but about the evolution of all possible histories (i.e. The Ruliad). And in most cases it won't have things to say about \emph{particular} histories. But instead what it will describe is what an observer sampling the whole multicomputational process will perceive, to which we now turn.

\begin{definition}
\emph{Multicomputation}, also known as the multicomputational paradigm, is a generalization of the traditional computational paradigm to encompass multiple computational histories or threads of time. In the standard computational approach, time progresses in a linear fashion. This means that the next state of a system is computed successively from its previous state. In contrast, multicomputation allows for every possible path of  computation to proceed through distinct, interwoven threads of time. Instead of a single linear progression, there are thus multiple threads of computational time that can be explored. In essence, multicomputation expands the scope of computational exploration by considering all possible computations simultaneously, rather than just one at a time. 
%\end{definition}

\vspace{0.2cm}
\noindent The following provides a characterization of how computational steps are executed in a multicomputational paradigm:
\vspace{0.2cm}

%\begin{definition}
\noindent Let $S$ be a system defined by a set of computational rules $R$ and initial conditions $I$.
In the traditional computational paradigm, the evolution of $S$ is represented by a sequence of states $s_1, s_2, \ldots, s_n$ such that each state $s_i$ is derived from $s_{i-1}$ using the rules $R$.
In the multicomputational paradigm, the evolution of $S$ is represented by a network $T$ of states, where each node represents a state of $S$ and each branch represents a possible path of computational evolution  based on  rules $R$. Each node can have multiple child nodes, representing different possible next states. Furthermore, state equivalences between nodes in different branches allows for intersections of different evolution paths. In other words, for each state $s_i$ in $T$, there exists a set of states $C\left(s_i\right)$ such that for each $s_j$ in $C\left(s_i\right), s_j$ is a possible next state of $s_i$ following rules $R$. The network  $T$ thus captures all possible computational trajectories of $S$ starting from initial conditions $I$.
\end{definition}

\begin{definition}
\emph{Ruliology}: Derived from the term ``rule,'' is the systematic study and exploration of computational rules and their myriad manifestations within computational systems. It delves into the intricacies of rule space, examining how diverse collections of rules can give rise to complex behaviors and structures. Ruliology transcends traditional computational boundaries, aiming to comprehend the foundational principles behind all possible computations, and seeking to understand how distinct collections of rules can generate entire universes of computation. At its core, Ruliology is an attempt to map out and understand the vast, multifaceted landscape of the Ruliad, where every conceivable rule is executed in every possible way.

\vspace{0.2cm}
\noindent In terms of the Ruliad, Ruliology as the study of computational rules and universes can be characterized as follows:
\vspace{0.2cm}

\noindent Let $R$ be the collection of all possible computational rules, $S$ be the collection of all possible  states, and $F: R \times S \rightarrow \mathcal{P}(S)$ be a map denoting the evolution or transformation of states dictated by a rule from $R$ upon a state from $S$.  The Ruliad $\mathbf{R}$ (defined above) includes all such computational evolutions.  Ruliology is then the study of the properties, structure, and implications of the Ruliad $\mathbf{R}$, as well as the exploration of individual and collective behaviors arising from elements of $R$ when acted upon $S$.
\end{definition}

%Hatem = Can you say what the other three paradigms are here???

\section{Observers, Sampling and the Physical World}

\begin{quote}
O God! I could be bounded in a nutshell, and count myself a King of infinite space...
William Shakespeare, \emph{Hamlet}, II, 2
\end{quote}

%Observers as introducing an interface between system and environment

\noindent A potentially serious stumbling block with the Ruliad idea as it stands is what we might call the ``realization problem'': how does an abstract rule get turned into physical reality? How do we end up with a particular history, to use the previous section's terminology? If this reality is the result of the computation of rules, then what is \emph{doing} the computation? Surely a computer of some kind? But then if this is a fundamental theory should this computer not itself be part of the Ruliad? We find ourselves in this way in a loop which cannot possibly be a virtuous circle of reasoning. It is more akin to that famous adventurer Baron von Munchausen rescuing himself and his horse from a quagmire by lifting himself up by his own hair. We can expect something like this problem to face any pre-geometry-type proposal that intends to dig beneath the spatiotemporal world populated with matter to something more abstract lying beneath---in several quantum gravity proposals the task of getting the world we are acquainted with, with its description in terms of fields on differentiable manifolds, from a deeper discrete theory, is known as the ``reconstruction problem''. Part of the problem is that the deeper theories do not involve things spatially located, evolving dynamically, but an abstract and far more primitive structure, often based on more relational concepts such as graphs and networks. It is, of course, a general and well-known problem to explain how we move from abstract formalism to physical reality. Usually, we start from the physical reality as a foundation, and then develop an abstract representation of it. In the case of pre-geometrical approaches (such as the Wolfram model), one makes no initial theoretical assumptions about the nature of physical reality, but starts instead from an abstract domain, with the hope of then recovering the physical aspects from this.

A related problem, of moving from abstract to concrete, is well expressed by John Wheeler \cite[p. 1208]{wheegrav}:

\begin{quote}
    Paper in white the floor of the room, and rule it off in one-foot squares. Down on 
one's hands and knees, write in the first square a set of equations conceived as able 
to govern the physics of the universe. Think more overnight. Next day put a better 
set of equations into square two. Invite one's most respected colleagues to contribute 
to other squares. At the end of these labors, one has worked oneself out into the doorway. 
Stand up, look back on all those equations, some perhaps more hopeful than others, 
raise one's finger commandingly, and give the order ``Fly!'' Not one of those equations 
will put on wings, take off, or fly. Yet the universe ``flies''.
\end{quote}

% XDA - Hawking has a related quote, added below

\noindent A well-known related sentiment was expressed through Stephen Hawking’s question ``What breathes fire into the equations?'' in his book, \emph{A Brief History of Time} \cite{SH}.  In other words, what make an (abstract) equation or generalization (which `oversees' a set of possibilities) a physical reality? Hawking elaborates:

\begin{quote}
Even if there is only one possible unified theory, it is just a set of rules and equations. What is it that breathes fire into the equations and makes a universe for them to describe? The usual approach of science of constructing a mathematical model cannot answer the questions of why there should be a universe for the model to describe. Why does the universe go to all the bother of existing?
\end{quote}

\noindent This is an issue concerning the metaphysics of the laws of nature. We can give much the same response for the Ruliad as Wheeler did here: observership, or rather participation, provides the necessary engine that powers the creation of the physical universe. Hence, the Ruliad will need to include a ``theory of the observer'' such that ``the universe as we know it'' seen from the vantage point of ``observers like us'' realizes observable physical attributes of the universe. However, it would be a mistake to view the physical universe as \emph{unique} and \emph{absolute}. Moreover, any computation is a result of the perspective of a computationally-bounded observer, rather than a fundamental feature of the universe itself: we are now in the realm of epistemology (i.e. description or representation) rather than ontology (how things are in a fundamental sense). The observer acts as a kind of transducer for the Ruliad, converting the abstract computations into (apparently) physical form.\footnote{Though ultimately everything (observer and observed) is supposed to remain part of the Ruliad of course, and so will remain abstract when conceived from a third-person perspective. However, if given a complete treatment, then the notion of the Ruliad is also representation via an observer.}

There is here an interesting philosophical relation of the Ruliad to Leibniz's system of monads. Leibniz's \emph{monadology}, his last attempt to codify his philosophical system, can certainly rival Wolfram's Ruliad for all encompassing majesty, despite its extreme brevity. Each monad is an individual that reflects the rest of the universe from its own unique point of view. The parts shape the whole and in turn, the whole back-reacts on the parts. Likewise, the Ruliad has similarity to \emph{Indra's Net} from \emph{The Flower Garland Sutra} - a kind of representation of a totality in terms of bejeweled vertices which encode the whole. Each is a vista of the whole. Every possible view is present in the whole. It is interesting to see how this basic idea, in which a totality is decomposed into an interdependent parts, repeats.\footnote{See \cite{atrick} for a treatment  of such a decompositional  metaphysical position (decompositional dual-aspect monism) as elucidated in several case studies from physics, of which the Wolfram model appears to be another convincing instance.  }

While monads collectively provide all possible perspectives of a world, as tiny independent mirrors (or points of view), Wolfram's Ruliad deals with all possible rules applied to some initial collection of abstract relations.  However, there is also a sense in which the monads are carrying out a pre-set program (or entelechy), coordinated with all other monads, in a pre-established and divinely choreographed dance determined to generate (i.e. construct) the best of all possible worlds. While there are the well-known principles of sufficient reason and identity of indiscernibles providing basic constraints on this construction, the principles themselves do not directly determine what is constructed. Rather, they inform the composition of the monads into complex structures which is then carried out through the pre-established harmony. A major reason for the introduction of  pre-established harmony was to explain the mind-body (or soul-body) correlations. For Leibniz there was no causal link and the correlation stemmed from the common cause in which both were set on their way like a pair perfectly synchronized watches. Interestingly, as we will develop further in another paper \cite{paper2}, Wolfram's model has a remarkably similar explanation for the correspondence of the world to the mind in that they both emerge from the same initial rules for construction and emerge in parallel with the mind (or observer) simply sampling the world and providing a perspective, much like a monad, where different observers represent the whole universe from different points of view. Likewise, one can find a similar generation of variety in the Wolfram model through this dislocation of a single, unified structure into many of points of view. 

Of course, Leibniz's theory, as it stands, is not one of much practical value in terms of showing how our present theories and phenomena can be \emph{constructed}. The approach of the Wolfram model, involving hypergraph rewriting systems,  places the ontological weight on the very rules of construction themselves. By contrast with Leibniz's ``God as  architect'' (as he puts it in S.89 of his \emph{Monadology}), here the metaphor is better expressed, following Chaitin \cite{chaitin}, as ``God as programmer,'' though here employing a multiway approach rather than a single-track, Turing machine approach. A more crucial distinction, related to the constructive approach, is that the physics (and mathematics) emerge from the interplay of computationally-bounded, embedded systems (observers) with the structure in which they are embedded (and therefore sampling), namely the Ruliad.

What an observer ultimately does is to take in an input from a large set, and return an output from a much smaller set, thus acting as a kind of idempotent filter. It's a concept that's appeared in many fields under many different names. It can be called a contractive mapping, a reduction to canonical form, a classifier, a forgetful functor, lossy compression, projection, renormalization group transformation, and so on.  It's what's fundamentally going on whenever we use a sensor or a measuring device, or for that matter, our human senses: we extract statistics, fit to models, and describe things symbolically. As a basic physical example, consider a gas pressure sensor based on a piston.  Within the gas, individual molecules move around in complicated and seemingly random ways, hitting the piston in all kinds of configurations.  But the piston ``reduces out'' all those details, responding just to the aggregate force of all the molecules, the same one of which can be realised in potentially infinitely many ways.  The main point is that we can describe what's going on more formally by saying that ``observations by the piston'' identify all the different detailed configurations of molecules, preserving only information about their aggregate force, forgetting the finer details.

This same idea can then be given slightly different interpretations, revealing how observers influence various branches of science. In statistical physics, for example, observers have, as just described, the effect of averaging over many particles or other degrees of freedom \cite{sw2law}. In General Relativity, they are averaging over spacetime regions, and forgetting those details having to do with coordinate transformations. In quantum physics, they are basically averaging over many quantum histories.  In mathematics the ``same'' statements are stated differently in terms of underlying axioms (see \cite{w3}). Gauge theories can be understood in the same way: the equivalence classes will in this case be generated by the gauge transformations which will be identified relative to an observer (though perhaps some other observer could view absolute structures such as the individual gauge potentials, much as a skilled musician with absolute pitch can hear differences that most others identify). In high energy physics,  black holes in various dimensions have an associated thermodynamic description, such that the physical charges of the black hole or black ring depend on whether the object being is viewed from a 4D perspective or a 5D perspective   \cite{5d1,5d2}.  Further afield, in economics, the focus might be on certain indices generated by the behaviours of a country's people. Even in linguistics we have identifications (between systems that differ in the details) given by the equivalence relation defining concepts, such as what counts as a chair, coat hook, or cabbage. This can then be applied to all scientific areas (and beyond) in which observers are involved.\footnote{Note that we are able to see a clear explanation of the so-called ``unreasonable effectiveness of mathematics'' here \cite{wigner}. The mathematical and physical structure emerge from the selfsame source, namely ruliadic sampling. Since they are constrained in the same way, a specific observer's sampling system will pick up correlations between the systems and properties it generates, and the way they are encoded in mathematics.}

Let us now consider some further conceptual implications of this overall structure or observers sampling the Ruliad, along with the notion of updating/rewrite rules. We start with the status of the approach \emph{vis-\`{a}-vis} determinism and modality.

\section{Computation,  Determinism and Free Will}

The question of determinism versus indeterminism is not so clear-cut in the Ruliad model and intersects with the issue of descriptions from the inside versus outside [i.e. computationally unbounded versus computationally bounded respectively]. One reason is that the inclusion of quantum mechanics into this picture is achieved through a notion of a branching (i.e. multiway), rather than linear, structure connecting the states in a process. This is supposed to represent the multiplicity of possible paths that quantum, but not classical, mechanics entails. In much the same way that the universal wave function, while itself deterministic (evolving according to the linear Schrodinger equation) nonetheless contains a kind of \emph{local} indeterminism if one follows specific paths through the space defined by the universal wave-function.\footnote{We have in mind the Deutsch-Wallace \cite{wal} approach to the interpretation of probabilities in quantum mechanics according to the Everett interpretation in which one also has to square an ultimately deterministic, branching process with the apparent indeterminism in measurement results. Indeed, there is no reason why one might not adopt the self-same decision-theoretic approach in the Wolfram model in which `things occur' (i.e. outcomes) only relative to the embedded observers. The idea would, in this case, be to view the probabilities in terms of rational decision under uncertainty about ones location in the Ruliad.} In other words, our answer to the question ``Is this theory deterministic or indeterministic?'' hinges on whether we are viewing things from an embedded perspective or from a God's eye perspective.

However, to view this branching itself as parallel to the indeterminism of quantum mechanics is to deny the possibility of classical, deterministic physics. And, of course, the Wolfram model should contain both classical and quantum theories, according to its role as a theory of all theories. Wolfram has a way of explaining this, of course, by pointing to a coarse-graining effect mentioned in the previous section.   The explanation of definite happenings [phenomena] despite the multiplicity of paths then comes about through the observer-participator as embedded in the multiway system. However, as with many-worlds interpretations, from the perspective of the totality (the Ruliad), everything that can happen (from the point of view of rules) does happen. However, the approach here goes beyond many-worlds---or is perhaps more in keeping with Hugh Everett's original ideas---since it incorporates the observer in the model itself (something we explore more in the next section). 

Given the branched-system understanding of quantum mechanics, it is clear that the model can support a grounding of counterfactuals along the same lines as constructor theory, in which the Everettian approach is adopted (at least for counterfactuals that occur in some branch). One of the problems faced in that approach is precisely that there are modal claims (e.g. those involving the very laws of physics themselves) that do \emph{not} occur in a branch of the multiverse. In this sense, the Ruliad has more modal breadth, since it encompasses entire theoretical structures.

Does this have anything to tell us about free will? One might, for example, take the idea of computational irreducibility, in which one must let a system  evolve to know its evolution, as some kind of free will proxy. However, this is simply unpredictability, not indeterminism. We find the same behaviour in chaotic systems, of course, where it is understood that the systems are perfectly deterministic albeit with agent-relative uncertainty about the development. It is why we simulate some systems, and in which the best we can do to know a future state is simply increase the performance power of the computer cranking through each iteration. So as observers watching a computationally irreducible process, we have uncertainty about the future, but the future is not uncertain \emph{simpliciter}. This must be the case if the Ruliad is seen to exist eternally, so that it is not a growing block structure. Such a degree of uncertainty does not imply free will. Even if there was pure randomness, it would not imply free will. Free will is the ability for the evolution of a system to fork in such a way that an agent has the ability to decide which path is chosen. That is, one could have isomorphic histories up to a point $t$ which diverge thereafter. There seems to be a confusion occurring between one's actions not being pre-determined and free will. But with the Ruliad we have neither of these situations, but simply an epistemic uncertainty about the future based on embedding an observer in a system with a kind of temporal gradient that emerges from the hypergraph rewriting process that the selfsame observer will be part of.

Wheeler famously asked ``Why the quantum?'' We have an answer here: the Ruliad (and its included samplers). And if we are then pushed to ask Why the Ruliad? Then we have an answer there too: there could not \emph{not} be the Ruliad, since it is a mathematically necessary object. But we can say a little more here. The quantum is not only a matter of the multiway picture. It is also a demand that the observer be included. Without this we have something akin to a space of pure potentiality, but in which nothing is made concrete or actual. There are no happenings without the consideration of a frame with respect to which something happens. This is very similar to the way in which there is no moon in quantum mechanics when no one is looking: there simply is no objective \emph{way} the world is in quantum mechanics, and likewise not in the Wolfram model either. Rather, there are all ways, which implies no way. Hence the need to introduce something like a Wheelerian observer-participator to select one such way the world can be, though with uncertainty as to which way that is. This notion of including the observer in so central a way suggests that the Wolfram model would benefit from ideas originating in 2nd-order cybernetics. Indeed, the Wolfram model might be a fine example of a naturally 2nd-order cybernetic system.

\section{The Algorithmic Nature of Observers}
%XDA 

Circling back to the dual relation between computation and measurement of an observable, one may think of measurement as an inference process under conditions of uncertainty. In other words, the process of observation (at least in the sense of making measurements of physical quantities) is the act of inferring approximate or coarse-grained causes of the outcomes that computations (around the observer) generate. Given the observer's own computational boundedness, it has to base these estimates on limited samples and restrict to causes that provide description only up to those levels of complexity that its own inferential engine can handle. This is where making equivalences and coarse-graining states of high complexity become relevant for any observer theory. This suggests an algorithmic nature of the observer as an inference engine within the Ruliad seeking states or computations with lowest complexity. 

Apart from the setting of the Ruliad, a theory of inference engines has been described in the context of cognitive neuroscience in terms of Karl Friston's ``Free Energy Principle,'' where a cognitive agent seeks to minimize its free energy either by performing actions upon the world or by changing its perceptions/representations of the state of the world based on new incoming data \cite{friston}. The free energy here is a complexity measure\footnote{See \cite{con4} for an overview of complexity measures related to cognition and consciousness.} whose minimization is associated to minimizing the ``surprise'' or uncertainty in the agent's representation of the world around it. More generally, for cognitive agents (both, biological and artificial), this minimization is achieved algorithmically using a hierarchical inference scheme based on feedback loops involving predictions and errors concerning the states of the world in comparison to the agent's own prior expectation \cite{con2,morpho}. 

Coming back to the Ruliad, the idea of an observer as an inference engine may be abstracted as a theory of sampling and measurement of low-complexity (or at least comparable to the observer itself) states within the Ruliad. Any complexity minimization principle akin to the free energy principle is in fact a second-order cybernetic construct. Presumably, this has to be included as a meta-rule upon the Ruliad.

%%DR - this sounds correct: can you say more? Maybe in the next section?

In contrast, Roger Penrose has famously defended the view that human consciousness is non-algorithmic \cite{penrose} (see also \cite{con1,con3}). Prima facie, if we are treating observers as an emergent feature of the Ruliad, then we must respond to Penrose's challenge. If we take Penrose's view of consciousness, as developed with Stuart Hameroff \cite{hp}, then we can see how this can be accommodated by accommodating the microtubles within the Ruliad, and having an account of the coherence they exhibit. 

Penrose and Hameroff posit that orchestrated objective reduction of the wave function is associated to proto-consciousness and this is non-computable. When Wolfram speaks of computation as omnipresent, he refers to a general use of the term that includes both computations that are reducible as well as irreducible. It is the irreducible ones that correspond with what Penrose refers to as non-computable. From the point of view of microtubuli represented within the Ruliad, they are running some irreducible rules (analogous to CA rule 30 \cite{nks}) which to another observer (within the Ruliad) does not lend itself to full predictability. When an observer conflates histories of the multiway, that constitutes a measurement upon the external world. This measurement itself may be computationally reducible. But the observer also needs a higher-order computation which determines which measurements to make and which histories of the multiway it should conflate - that higher-order process may be irreducible (one may call that meta-cognition). If these higher-order processes are required for consciousness, then the conscious observer is not just a program, but a meta-program (and an irreducible one). 

Ultimately, what computational irreducibility means is, as the name suggests, that there is no redundancy in the process that can be eliminated to shorted or compress the process. This means that there can be no `short cuts' in which features of the process can be ignored, made equivalent, or in some other such way utilised to jump to the end. The best one can do is to run the process or simulate it. Again, one might be able to throw more performance power at at, but still it must run through step-by-step. In this sense we see that predictability is bound up with the notion of reducibility, and we have something like an open future if not quite full-blown free will. We are thus left with incompleteness, however, which is related to the so-called hard problem of consciousness. We are giving a model of an observer `from the outside' as it were. Yet how do we find a place for subjectivity (the inside view) here?

% XDA

\section{Seeing the Ruliad from the Inside: Second-Order Cybernetics}

\begin{quote}
\noindent Space and time, defining everything we cognize by sensuous means, are in themselves just forms of our receptivity, categories of our intellect, the prism through which we regard the world - or in other words, space and time do not represent properties of the world, but just properties of our knowledge of the world gained through our sensuous organism. From this it follows that the world apart from our knowledge of it, has neither extension in space nor existence in time; these are properties which we add to it. (P. D. Ouspensky, \emph{Tertium Organum}, 4)
\end{quote}

\noindent The notion that the world we experience (the phenomenal world or manifest reality) is \emph{conditioned} by our faculties as observers, including the notion of computational boundedness or limitation, can be traced to Kant's theory of the categories. This traces many features that we might naively impute to the world itself back to features of the observer. A natural question, and one considered by Kant, is what happens when different observers are considered. The Wolfram model also involves the idea that different observers might generate very different descriptions of the Ruliad, and so would discover different laws in their world. It is, in other words, vital that the specifics of observers be provided, in order to get a world-description out, and as such the former is the \emph{sine qua non} of the latter.

Several examples of such `alien' scenarios were presented in the early flatland ideas. While Edwin Abott’s approach is the best known, the most useful for our purposes is Charles Howard Hinton’s, who writes:

\begin{quotation}
\noindent Thus if we make up the appearances which would
present themselves to a being subject to a limitation or
condition, we shall find that this limitation or condition,
when unrecognized by him, presents itself as a general
law of his outward world, or as properties and qualities
of the objects external to him. He will, moreover, find
certain operations possible, others impossible, and the
boundary line between the possible and impossible will
depend quite as much on the conditions under which he
is as on the nature of the operations. \cite[p. 40]{hinton}
\end{quotation}

\noindent Our epistemological equipment allows us to generate a kind of screen on which reality can display itself. But, of course, what is manifest is only a relative appearance and has much to do with the equipment (including any necessary factors that enable it to exist in the first place). Thus, the observer (human or otherwise) acts as a kind of prism, or transducer, converting a potentially infinite spectrum of data into a finite package capable of being processed. The prism is a good analogy because without it, there would be no such phenomena. And had we placed a distinct observer where the prism is, perhaps a mirror, then we would generate a very different kind of display of the \emph{same} region.\footnote{Note also that some of these `displays' might be mutually incompatible leading to the kinds of complementarity one finds in quantum mechanics, e.g. with the inability to place equipment capable of both position and momentum measurements.} But, to repeat, without some means of displaying the world, there is nothing there other than a kind of potentiality to display.

Second-order cybernetics is based on the idea that no science is possible from a `view from nowhere' in which one can view reality unveiled as it were \cite{cyb1,cyb2,cyb3,vfUU}. One has to consider a standpoint, or perspective, or frame from which the universe is viewed. Without it (i.e. the viewer), there is no view. The Ruliad as it stands, is abstractly defined as a view from nowhere: a totality. Wolfram himself speaks explicitly of the Ruliad ``viewed from the outside'' \cite[p. 235]{w1}. To carry out scientific exploration in the Ruliad, we must include a system, an observer, capable of sampling the space.\footnote{We don't go into any details here, beyond simply noting that the Wolfram model fits the basic mould of the 2nd-order cybernetic framework. A future paper will consider the pairing in more details. See \cite{vf} for a superb review of the basic ideas of 2nd-order cybernetics.}

Wolfram elsewhere defines the Ruliad as ``result of following all possible computational rules in all possible ways'' \cite{ruliad,w3}. This is more in line with the 2nd-order cybernetics approach, but we must ask: \emph{who} is following the results? Who is the observer in this case? And who models that observer? From the outside, the Ruliad is simply understood as the totality of all possible computations. However, from the point of view of any of its parts (which satisfy criteria of computational boundedness and  persistence), any part of the Ruliad, qualified by boundedness and persistence, is potentially an observer of its complement (within a specified horizon, which would again depend on its computational boundedness). The computations performed by this localized observer realize measurements in the universe. Hence, from the perspective of 2nd-order cybernetics, the object of  interest may not be the Ruliad by itself, but rather the power set (or appropriate categorical generalization of a power set) of the Ruliad that encapsulates observers, the observed and their interactions. Teleology and mereology will both be relevant to this power object (of the Ruliad). 
%XDA

%%% HE %%%

In the past, it was usually possible to do theoretical science without explicitly discussing the observer. But it turns out that to say anything about ``what happens'' requires knowing about the observer. In general, what an observer does is take the raw complexity of the world, and reduce it in such a way that conclusions can be made from it.  And here one can think of a certain fundamental duality between computation and observation: the process of computation has the effect of generating new outcomes; the process of observation has the effect of reducing outcomes by `equivalencing' different ones together, as discussed earlier. Computation theory gives us a way to describe possible processes of computation.  And the goal of what we're calling ``observer theory'' is to give us an analogous way to describe possible processes of observation. Because it turns out that our limitations as observers are in a sense what gives us many of the most fundamental scientific laws that we perceive. And it's really all about the interplay between the underlying computational irreducibility and our nature as computationally bounded observers.

The crucial feature of observers seems to be that the observer is always ultimately some kind of ``finite mind'' that takes all the complexity of the world and extracts from it just certain ``summary features'' that are relevant to the ``decisions'' it has to make \cite{swcon}. Observers like us have two basic characteristics: first, that they are computationally bounded, and second, that they are persistent in time. Computational boundedness is essentially the statement that the region of space observers occupy is limited, i.e. we can't expect to ``reverse engineer'' computationally irreducible processes that are going on ``underneath'' \cite{swcon}.  

The Wolfram model takes the perspective that an observer has to be a part of the underlying multiway system (possibly as a subgraph spread across branches). In this view, measurement is consequently the process of the observer conflating parallel threads of multiway history with a single evolution leading to the illusion of a unique sequential thread of time. Furthermore, the notion of causal invariance, which can be thought of as being associated with paths of history that diverge eventually converging again, is what guarantees a coherent eventual consistency. And since the Ruliad contains paths corresponding to all possible rules, it's basically inevitable that it will contain what's needed to undo whatever divergence occurs---because of causal invariance the laws of physics are invariant in any frame of reference (though to realize the laws one has to set up a system within a given reference frame). Hence, anything physically observable is going to be in a subjective setting, and from this subjective setting, when we infer anything objective, it is relative-objectivism. 

So any given observer is interpreting what they see in terms of a description language, which causes them to attribute certain rules to be ``the rules of the universe''---one has to choose what kind of system one is working with, and it is almost impossible to state a law without those choices. Once we make those choices, we are already in a constructive domain. Hence, if one sets up some particular computational system or mathematical theory there will always be choices to be made, and our most important feature as observers is that we're computationally bounded, i.e. the way we parse the universe involves doing an amount of computation that's absolutely tiny compared to all the computation going on in the universe. We sample only a tiny part of what's really going on underneath, and we aggregate many details to get the summary that represents our perception of the universe. Recall our earlier example of the molecules in a gas. The molecules bounce around in a complicated pattern that depends on their detailed properties, but an observer like us doesn't trace this whole pattern. Instead, we only observe certain ``coarse-grained'' features (e.g. pressure and temperature). In this sense everything then boils down to how an observer chooses/samples the space in which they are located, so that their properties are of the essence, which reveals  the Wolfram framework as an already 2nd-order cybernetic system.

\section{The Limits of Ruliology: The Impossibility of Seeing the Ruliad from the Outside}

\begin{quote}
    Philosophy is an attempt to express the infinity of the universe in terms of the limitations of language.

A. N. Whitehead

\end{quote}

\noindent If, as Whitehead put it, philosophy is an attempt to express the infinity of the universe in terms of the limitations of language, Ruliology is likewise an attempt to express that infinity in the somewhat less limited framework of the representations of computationally-bounded observers that are embedded within it. The Wolfram model depends on coarse-graining over paths in order to model the observed physics. The coarse-graining is, of course, relationally linked to specific observers (or classes of observers). This introduces a limit, since any ruliadic properties that we can speak about are of course from the point of view of a member of such a class of observers. We can model other observers by changing the properties defining the observer-class, but even this  is itself generated from our own perspective and so will inherit any associated limitations.

Following on from the two ways of thinking about the Ruliad, from the inside versus the outside, we can see that the irreducibility is a feature of the embedded view: it is a feature of the relationship between observers and the Ruliad of which they are a part.  We can link this to the two broad approaches to the ontology of mathematics, and note the role they play in physics. From the inside view, in this case, the appropriate ontological picture is that of constructivism, with intuitionistic logic playing the role, and in someway paralleling the computational irreducibility that the observers face in their knowledge claims. But taken as a completed object, where all processes have been carried out to their infinite limits, then there is of course no computational irreducibility, because the object is eternally given, and we view it \emph{sub specie aeternitatis}.

It seems that Wolfram is acutely aware of the necessity to include the observer in the description itself \cite{sw2020}:

\begin{quotation}
    It's a typical first instinct in thinking about doing science: you imagine doing an experiment on a system, but you---as the ``observer''---are outside the system. Of course if you’re thinking about modeling the whole universe and everything in it, this isn't ultimately a reasonable way to think about things. Because the ``observer'' is inevitably part of the universe, and so has to be modeled just like everything else.\footnote{\url{https://writings.stephenwolfram.com/2020/04/finally-we-may-have-a-path-to-the-fundamental-theory-of-physics-and-its-beautiful/}}
\end{quotation}

\noindent 

Wolfram writes that \cite{w3}:

\begin{quotation}
    [T]he Ruliad is not just a representation. It’s in some way something lower level. It's the ``actual stuff'' that everything is made of. And what defines our particular experience of physics or of mathematics is the particular samples we as observers take of what’s in the Ruliad.\footnote{\url{https://www.wolframscience.com/metamathematics/counting-the-emes-of-mathematics-and-physics/}}
\end{quotation}

\noindent This stuff is made of ``emes,'' which function as the most fundamental layer \cite{w3}. It is supposed to transcend the observers, and goes beyond representation. However, given our discussion of 2nd-order cybernetics, can this be right? How can we possibly make any statements about reality that do not carry with them their source from us \emph{qua} observers? To say it is not just representation implies that we can somehow step outside of all representation, and step outside of our position as observers, to see that more lies beyond: \emph{plus ultra}. In doing so we have stepped outside of ruliology proper, and entered speculative metaphysics. Rather, as Heinz von Foerster puts it:

\begin{quotation}
    [A] brain is required to write a theory of a brain. From this follows that a theory of the brain, that has any aspirations for completeness, has to account for the writing of this theory. And even more fascinating, the writer of this theory has to account for her or himself. Translated into the domain of cybernetics; the cybernetician, by entering his own domain, has to account for his or her own activity. \cite[p. 289]{vfUU}
\end{quotation}

\noindent The observer in the Wolfram model must ultimately also be ruliadic if this theory is to be truly fundamental.  Indeed, the necessity for second-order cybernetics suggests the need for \emph{meta-rules} upon the Ruliad itself. These rules when instantiated locally within computationally bounded patches of the Ruliad operationalize an abstract notion of observers.

In this case, when we speak of ourselves sampling the Ruliad to generate particular systems of mathematics and physics, we are really speaking about the Ruliad \emph{self}-sampling.\footnote{This is also known as endophysics, or the physics from the inside (see e.g. \cite{rosser}). R\"{o}ssler argues that the world is always relative to an observer-perspective, so that an ``interface'' is involved in which the world appears as a kind of screen to the observer, though not in any fixed way. Rather, the cut between self and world (or observer and observed/environment) is  variable. In the case of the Observer-Ruliad system the observer is a kind of bounded foliation of the whole. } In this way we can compare the role and status of observers in the world to the role of humans in such religious systems as Sufism. A totality has splintered into many (relative) perspectives, each with the ability to explore a particular part of that totality.  The perspectives are transducing the ineffable Ruliad (akin to an \emph{absolute} unconditioned reality) into something `effable.'  The self-sampling naturally leads to reflexivity and looping elements, linking the observer and observed. As Kauffman explains:

\begin{quotation}
   \noindent  In an observing system, what is observed is not distinct from the system itself, nor can one make a separation between the observer and the observed. These stand together in a coalescence of perception. From the stance of the observing system all objects are non-local, depending upon the presence of the system as a whole. It is within that paradigm that these models begin to live, act and converse with us. We are the models. Map and territory are conjoined. \cite[p. 1]{kauf1}
\end{quotation}

\noindent This is not a flaw with such a model, but rather a virtue. It is quite clear that if we consider some global system [a universe or Ruliad] then we can see that is is trivially the case that some sub-system [the observer] of that system cannot observe the whole system. In other words, the system as a whole is not an observable in the strict sense: there is no operation that we can envisage to measure it.\footnote{Here we are assuming that the system is closed, of course, so that there is no external interaction and no sense in which the system is itself a sub-system of some larger system. In the case in which, e.g., the system is a simulated universe, then of course one can well imagine an external observer being able to make appropriate measurements on the system, though they would then face the same problem in their own universe (cf. \citep[2.1]{fields}, \cite{fields0}) and the description in question would no longer be fundamental.} That the Wolfram model contains  observers and their viewpoints, and generates physics through sampling [in a consistent  loop], is the most fundamental model one can manage if one does indeed take seriously the fact that we ourselves must be such observers. It is a virtue for a theory to describe its limitations, in this case containing its limits as theoretical outcomes.

Of course, the Ruliad is not something that we could ever directly observe, and nor is it presented as such. It is an abstract entity that if picturable in any way would be akin to a kind of hyper-tensorial object. However, inasmuch as it is abstract, it exists as representation in the mind of an agent and so inherits the limitations. In this case we cannot quite speak of the distinction between map and territory blurring, as with perceptions of the world, because the Ruliad is supposed to go beyond any possible perception. In this sense it stands more in the position of an unknowable God, and the evidence we have is more of the form of a transcendental argument, such that it functions as the \emph{ground} required for having the kinds of experiences we do have and for there being an apparently existing universe in the first place. It is the ground of the possibility of multi-computation.

%purusha to jivatma?

As we are familiar with from general relativity, although there are many systems of coordinates that can serve to fix a gauge on the universe, each providing a foliation of the spacetime that is invariant when completed. There is also the notion of the quotient space that, in a sense, averages over all the gauge freedom, spitting out the invariant structure.\footnote{In a similar way, Wolfram speaks of ``bulk'' features, in the sense of something like the quotient object, i.e. that which transcends the baggage brought by observers and their gauges or frames.} Of course, the resulting entity is harder to deal with from a physical point of view since the gauge (the coordinate frame) are what allows us to epistemically access the universe. Of course, this also tells us that the epistemic access is also partly one of construction of what is observed. The frames are observer-constructs and must be purged in any consideration of what is the `true picture' of reality. Of course, one can also simply speak of the frame-relative picture in such a way that so long as the frame is considered in the evaluation of some physical quantity there is a kind of observer-invariance of the quantities by virtue of observers being able to change their own coordinates into new frames.

This issue of attempting to describe a reality beyond our selective, descriptive capacities is a common problem facing those dealing with apophatic theology, in which one cannot speak of the thing in question with positive characteristics because they thereby bound it, and yet the very bounds come from us. But without such an extremely deep level of probing, we cannot be said to have a fundamental theory. By focusing on the building rules themselves we find both physics and mathematics emerging, which is as we should expect.  Along these lines, a complaint that has been levelled against the Wolfram model described here is that it is incapable of making predictions. The same complaint was lodged against Eddington's Fundamental Theory. But it entirely misses the point, which is that the Ruliad is a home for physical theories. It is the ground. 

A fundamental theory, that is this fundamental, cannot possibly be expected to make direct predictions of the sort that the critics clearly desire. But what it can do is locate them in a web of theories, and moreover it can suggest entirely new kinds of theory that would then themselves make predictions when properly worked out in the manner appropriate for less fundamental approaches. The task of physics, indeed, is to figure out where in the Ruliad we are located. In this sense physics (and mathematics) amount to the task of a librarian working in Borges' Total Library. Every possible book is contained within that library, much as every possible theory is contained in the Ruliad. While all possible books are therein, one needs the right indexing system to locate the correct book, so as not to grab some book of gibberish. The observer is the linchpin that connects the Ruliad with scientific theories, since it is the locus of indexing.  Ruliology is not then, a replacement of physics, but a way of making sense of it. Moreover, what might appear to be physically nonsensical sectors of rulial space to us, might be perfectly experienciable (as a quite different kind of universe) to other kinds of observers. What we have described is, then, not a theory of physics in the ordinary at all. It is a pre-physical framework for any possible theory of physics and should not be analysed (or critiqued) in the same terms as orthodox physical theories. 

\vspace{.7cm}

\subsection*{Acknowledgments}
We thank Stephen Wolfram for engaging discussions on the Ruliad. This work was supported by the following grants:  The Foundational Questions Institute and Fetzer-Franklin Fund, a donor advised fund of Silicon Valley Community Foundation [FQXi-RFP-1817]; The John Templeton Foundation [Grant ID 62106]; and the Australian Research Council [Grant DP210100919].

\vspace{.7cm}




\begin{thebibliography}{99}

\bibitem{cqm1}
Abramsky, S., Coecke, B. (2009). Categorical quantum mechanics. \emph{Handbook of quantum logic and quantum structures}, 2, 261-325.

\bibitem{5d1}
Arsiwalla, X. D. (2008). More Rings to Rule Them All: Fragmentation, 4D $\leftrightarrow$ 5D and Split-Spectral Flows. \emph{Journal of High Energy Physics}, 2008(02), 066.

\bibitem{5d2}
Arsiwalla, X. D. (2009). Entropy Functions with 5D Chern-Simons Terms. \emph{Journal of High Energy Physics}, 2009(09), 059.


\bibitem{cpost}
Arsiwalla, X. D. (2020). Homotopic Foundations of Wolfram Models. \emph{Wolfram Community}. \url{https://community. wolfram. com/groups/-/m, 2032113}.


\bibitem{opalg}
Arsiwalla, X. D., Chester, D., Kauffman, L. H. (2022). On the Operator Origins of Classical and Quantum Wave Functions. \emph{ArXiv:2211.01838}.

\bibitem{paper2}
Arsiwalla, X.D., Elshatlawy, H., Rickles, D. (forthcoming). Pregeometry from Formal Language. \emph{In preparation}.

\bibitem{pregeo}
Arsiwalla, X. D.,  Gorard, J. (2021). Pregeometric Spaces from Wolfram Model Rewriting Systems as Homotopy Types.  \emph{ArXiv:2111.03460}.

\bibitem{nfold}
Arsiwalla, X. D., Gorard, J., Elshatlawy, H. (2022). Homotopies in Multiway (Nondeterministic) Rewriting Systems as $n$-Fold Categories. \emph{Complex Systems}, \textbf{31}(3), 261-277, \emph{ArXiv:2105.10822}.

\bibitem{con1}
Arsiwalla, X. D., Herreros, I., Moulin-Frier, C., Sanchez, M.,  Verschure, P. (2016). Is consciousness a control process?. In \emph{Artificial Intelligence Research and Development} (pp. 233-238). IOS Press.

\bibitem{con2}
Arsiwalla, X. D., Herreros, I.,  Verschure, P. (2016). On three categories of conscious machines. In  \emph{Biomimetic and Biohybrid Systems: 5th International Conference, Living Machines 2016, Edinburgh, UK}. Proceedings 5 (pp. 389-392). Springer International Publishing.

\bibitem{con3}
Arsiwalla, X. D., Signorelli, C. M., Puigbo, J. Y., Freire, I. T.,  Verschure, P. F. (2018). Are brains computers, emulators or simulators?. In \emph{Biomimetic and Biohybrid Systems: 7th International Conference, Living Machines 2018, Paris, France}, Proceedings 7 (pp. 11-15). Springer International Publishing.

\bibitem{morpho}
Arsiwalla, X. D., Solé, R., Moulin-Frier, C., Herreros, I., Sánchez-Fibla, M.,  Verschure, P. (2023). The Morphospace of Consciousness: Three Kinds of Complexity for Minds and Machines. \emph{NeuroSci}, \textbf{4}(2), 79-102.


\bibitem{con4}
Arsiwalla, X. D.,  Verschure, P. (2018). Measuring the complexity of consciousness. \emph{Frontiers in Neuroscience}, \textbf{12}, 424.


\bibitem{atrick}
Atmanspacher, H. and D. Rickles (2022) \emph{Dual-Aspect Monism and the Deep Structure of Meaning}. Routledge.

\bibitem{rewrite}
Baader, F., Nipkow, T. (1998). \emph{Term rewriting and all that}. Cambridge university press.


\bibitem{baez}
Baez, J. (2007). The homotopy hypothesis. Available online at \url{https://math.ucr.edu/home/baez/homotopy/homotopy.pdf}. 


\bibitem{chaitin}
Chaitin, G. (2018) Building the World Out of Information and Computation: Is God a Programmer, Not a Mathematician? In Wuppuluri Shyam \& Francisco Antonio Dorio (eds.), \emph{The Map and the Territory: Exploring the Foundations of Science, Thought and Reality} (pp. 431-438). Springer.


\bibitem{coeckevol}
Coecke, B., ed. (2011) \emph{New Structures for Physics}. Springer.


\bibitem{zx0}
Coecke, B., Duncan, R. (2011). Interacting quantum observables: categorical algebra and diagrammatics. \emph{New Journal of Physics}, 13(4), 043016.


\bibitem{cronin}
Cronin, L. and S. Imari Walker (2016) Beyond prebiotic chemistry: What dynamic network properties allow the emergence of life? \emph{Science} \textbf{352}(6290): 1174-1175.

\bibitem{constructor}
Deutsch, D. and C. Marletto (2015) Constructor theory of information. \emph{Proc. R. Soc. A.} \textbf{471}.

\bibitem{topos1}
D\"{o}ring, A.,  Isham, C. (2010). “What is a thing?”: Topos theory in the foundations of physics. In \emph{New structures for physics}, (pp. 753-937),   Springer Berlin Heidelberg.


\bibitem{graphc}
Du Plessis, J. F., Arsiwalla, X. D. (2023). A cosine rule-based discrete sectional curvature for graphs. \emph{Journal of Complex Networks}, \textbf{11}(4), cnad022.

\bibitem{einstein}
Einstein, A. (1936) Physics and Reality. In his \emph{Ideas and Opinions} (pp. 290–323). New York: Bonanza, 1954.

\bibitem{fields0}
Fields, C. (2012). If physics is an information science, what is an observer?. \emph{Information}, 3(1), 92-123.

\bibitem{fields}
Fields, C. (2016) Building the Observer into the System: Toward a Realistic Description of Human Interaction with the World. \emph{Systems} \textbf{4}(4): 32.

\bibitem{frauch18} 
Frauchiger, D. and R. Renner (2018) Quantum theory cannot consistently describe the use of itself. \emph{Nat Commun} \textbf{9}: 3711.

\bibitem{friston}
Friston, K. (2010)  The free-energy principle: a unified brain theory? \emph{Nature Reviews Neuroscience} textbf{11}: 127--138.

\bibitem{gisin}
Gisin, N. (2021)  Indeterminism in physics and intuitionistic mathematics. \emph{Synthese}, \textbf{199}: 13345–13371.

\bibitem{cyb1}
Glanville, R. (2002). Second order cybernetics. \emph{Systems Science and Cybernetics}, \textbf{3}, 59-85.

\bibitem{zx1}
Gorard, J., Namuduri, M., Arsiwalla, X. D. (2020). ZX-calculus and extended hypergraph rewriting systems I: A multiway approach to categorical quantum information theory. \emph{ArXiv:2010.02752}.

\bibitem{zx2}
Gorard, J., Namuduri, M., Arsiwalla, X. D. (2021). ZX-calculus and extended wolfram model systems II: fast diagrammatic reasoning with an application to quantum circuit simplification. \emph{ArXiv:2103.15820}.

\bibitem{zx3}
Gorard, J., Namuduri, M.,  Arsiwalla, X. D. (2021). Fast Automated Reasoning over String Diagrams using Multiway Causal Structure. \emph{ArXiv:2105.04057}.


\bibitem{hp}
Hameroff S. and R. Penrose (2014) Consciousness in the universe: a review of the 'Orch OR' theory. \emph{Phys Life Rev.}  \textbf{11}(1):39-78.

\bibitem{SH}
Hawking, S. (2009). \emph{A brief history of time: from big bang to black holes}. Random House.

\bibitem{cyb2}
Heylighen, F.,  Joslyn, C. (2001). Cybernetics and second-order cybernetics. \emph{Encyclopedia of physical science \& technology}, \textbf{4}, 155-170.


\bibitem{hinton}
Hinton, C. H. (1900) \emph{A New Era of Thought}. Swan Sonnenschein \& Co.

\bibitem{kauf1}
Kauffman, L. H. (2003) Eigenforms — Objects as Tokens for Eigenbehaviors. \emph{Cybernetics And Human Knowing} \textbf{10}(3-4).

\bibitem{marlettobook}
Marletto, C. (2021) \emph{The Science of Can and Can't: A Physicist's Journey Through the Land of Counterfactuals}. Penguin.

\bibitem{wheegrav}
Misner, C., K. Thorne, and J. Wheeler (1973) \emph{Gravitation}. Princeton University Press.

\bibitem{penrose}
Penrose, R. (1989) \emph{The Emperor's New Mind}. Oxford: Oxford University Press.

\bibitem{rick1}
Rickles, D. (2006) Time and Structure in Canonical Gravity. In D. Rickles, S. French and J. Saatsi (eds.), \emph{The Structural Foundations of Quantum Gravity} (pp. 152--191). Oxford: Oxford University Press.

\bibitem{hott3}
Riehl, E.,  Shulman, M. (2017). A type theory for synthetic $\infty $-categories. \emph{ArXiv:1705.07442}.


\bibitem{rosser}
R\"{o}ssler, O. E. (1987) Endophysics. In J. L. Casti and A. Karlquist (eds.), \emph{Real Brains, Artificial Minds}. North-Holland, New York.

\bibitem{ggrammar}
Rozenberg, G. (Ed.). (1997). \emph{Handbook of graph grammars and computing by graph transformation}, (Vol. 1). World scientific.


\bibitem{mod}
Schlichenmaier, M. (2007) \emph{An Introduction to Riemann Surfaces, Algebraic Curves and Moduli Spaces}. Berlin: Springer.


\bibitem{sch1}
Schreiber, U. (2013). Differential cohomology in a cohesive infinity-topos. \emph{ArXiv:1310.7930}.

\bibitem{sch2}
Schreiber, U. (2021). Higher Prequantum Geometry. \emph{New Spaces in Physics: Formal and Conceptual Reflections}, 2, 202.

\bibitem{sch3}
Schreiber, U.,  Shulman, M. (2014). Quantum gauge field theory in cohesive homotopy type theory. \emph{ArXiv:1408.0054}.

\bibitem{cyb3}
Scott, B. (2004). Second‐order cybernetics: an historical introduction.  \emph{Kybernetes}, \textbf{33(9/10)}, 1365-1378.


\bibitem{hott1}
Shulman, M. (2017). Homotopy type theory: A synthetic approach to higher equalities. In \emph{Categories for the working philosopher}, 36-57.

\bibitem{hott2}
Shulman, M. (2021). Homotopy type theory: the logic of space. In \emph{New Spaces in Mathematics}, (Vol. 1, pp. 322-404). Cambridge University Press.



\bibitem{hott4}
The Univalent Foundations Program. (2013). Homotopy type theory: Univalent foundations of mathematics. \emph{ArXiv:1308.0729}.


\bibitem{vfUU}
 von Foerster, H. (2003) \emph{Understanding Understanding: Essays on Cybernetics and Cognition}. New York: Springer.

 \bibitem{vf}
  von Foerster, H. (2014) \emph{The Beginning of Heaven and Earth has no Name: Seven Days with Second-Order Cybernetics}. Fordham University Press.

 \bibitem{wal}
 Wallace, D. (2003). Everettian Rationality: Defending Deutsch’s approach to Probability in the Everett Interpretation. \emph{Studies in the History and Philosophy of Modern Physics} \textbf{34}: 415--439.

\bibitem{wal2}
Wallace, D. (2012). \emph{The emergent multiverse: Quantum theory according to the Everett interpretation}. Oxford University Press, USA.

\bibitem{wigner}
Wigner, E. P. (1960). The unreasonable effectiveness of mathematics in the natural sciences. \emph{Commun. Pure Appl. Math. XIII}, 1-14. 


\bibitem{nks}
Wolfram, S. (2002). \emph{A New Kind of Science}. Wolfram media.

\bibitem{w1}
Wolfram, S. (2020). A Class of Models with the Potential to Represent Fundamental Physics.  \emph{Complex Systems}, \textbf{29}(2).

\bibitem{sw2020}
Wolfram, S. (2020). Finally We May Have a Path to the Fundamental Theory of Physics... and It's Beautiful.  \\    \url{https://writings.stephenwolfram.com/2020/04/finally-we-may-have-a-path-to-the-fundamental-theory-of-physics-and-its-beautiful/}. 

\bibitem{w2}
Wolfram, S. (2021) \emph{Combinators: A Centennial View}. Wolfram Media.

\bibitem{w3}
Wolfram, S. (2022) \emph{Metamathematics: Foundations \& Physicalization}. Wolfram Media.


\bibitem{ruliad}
Wolfram, S. (2021).  The Concept of the Ruliad. \url{https://writings.stephenwolfram.com/2021/11/the-concept-of-the-ruliad/}. 

\bibitem{swtm}
Wolfram, S. (2021). Exploring Rulial Space: The Case of Turing Machines. \emph{ArXiv:2101.10907}.


\bibitem{mcomp}
Wolfram, S. (2021).  Multicomputation: A Fourth Paradigm for Theoretical Science.  \url{https://writings.stephenwolfram.com/2021/09/multicomputation-a-fourth-paradigm-for-theoretical-science/}.

\bibitem{swcon}
Wolfram, S. (2021).  What Is Consciousness? Some New Perspectives from Our Physics Project. \url{https://writings.stephenwolfram.com/2021/03/what-is-consciousness-some-new-perspectives-from-our-physics-project/}.


\bibitem{sw2law}
Wolfram, S. (2023).  Computational Foundations for the Second Law of Thermodynamics. \url{https://writings.stephenwolfram.com/2023/02/computational-foundations-for-the-second-law-of-thermodynamics/}.


\bibitem{arity1}
Zapata-Carratala, C., Arsiwalla, X. D. (2022). An Invitation to Higher Arity Science. \emph{ArXiv:2201.09738}. In press \emph{Complex Systems}.

\bibitem{arity2}
Zapata-Carratala, C., Arsiwalla, X. D., Beynon, T. (2022). Heaps of Fish: Arrays, Generalized Associativity and Heapoids. \emph{ArXiv:2205.05456}.

\bibitem{arity3}
Zapata-Carratalá, C., Schich, M., Beynon, T., Arsiwalla, X. D. (2023). Beyond Binary: Hypermatrix Algebra and Irreducible Arity in Higher-Order Systems. \emph{ArXiv:2301.07494}.  
%Accepted in \emph{Advances in Complex Systems}. 

 
\end{thebibliography}


\end{document}